\documentclass[aps,prl,twocolumn,groupedaddress,draft,showpacs]{revtex4}
\usepackage{graphicx,epsf}
\newif\ifdraft
\drafttrue
\newif\ifpreprint
\preprinttrue

\def\fig#1{fig.~{\ref{#1}}}

\def\spa#1.#2{\left\langle#1\,#2\right\rangle}
\def\spb#1.#2{\left[#1\,#2\right]}

\def\Tr{\, {\rm Tr}}
\def\tr{\, {\rm tr}}
\def\SYM{MSYM}

\def\eqn#1{eq.~(\ref{#1})}

\def\eqns#1#2{eqs.~(\ref{#1}) and~(\ref{#2})}

\def\NeqFour{{\cal N}=4}

\def\be{\begin{equation}}
\def\ee{\end{equation}}
\def\bea{\begin{eqnarray}}
\def\eea{\end{eqnarray}}
\def\ba{\begin{eqnarray}}
\def\ea{\end{eqnarray}}

\def\gm{\gamma}
\def\Gm{\Gamma}

\def\ep{\epsilon}
\def\e{\epsilon}
\def\vareps{\varepsilon}

\def\lm{\lambda}

\def\dd{\mbox{d}}

\def\nn{\nonumber}

\def\oneloop{{(1)}}

\def\twoloop{{(2)}}

\def\Ord{{\cal O}}
\def\MB{{\tt MB\/}}
\newbox\charbox
\newbox\slabox
\def\s#1{{      
        \setbox\charbox=\hbox{$#1$}
        \setbox\slabox=\hbox{$/$}
        \dimen\charbox=\ht\slabox
        \advance\dimen\charbox by -\dp\slabox
        \advance\dimen\charbox by -\ht\charbox
        \advance\dimen\charbox by \dp\charbox
        \divide\dimen\charbox by 2
        \raise-\dimen\charbox\hbox to \wd\charbox{\hss/\hss}
        \llap{$#1$} }}
\def\ksl{\s{k}}

\begin{document}

\ifpreprint
UCLA/06/TEP/09
\hfill $\null\hskip 1.2 cm \null$  Saclay-SPhT--T06/032
\hfill $\null\hskip 1.2 cm \null$  WUE-ITP-2006-003
\fi

\title{Two-Loop Iteration of Five-Point ${\cal N}=4$ Super-Yang-Mills
Amplitudes}

\author{Z.~Bern${}^a$, M.~Czakon${}^{b,c}$, D.A.~Kosower${}^d$, 
R.~Roiban${}^e$ and V.A.~Smirnov${}^f$}
\affiliation{
${}^a$Department of Physics and Astronomy, UCLA, Los Angeles, CA
90095-1547, USA  \\
${}^b$Institut f\"ur Theor.~Physik und Astrophysik,
     Universit\"at W\"urzburg,
     Am Hubland, D-97074, Germany\\
${}^c$Institute of Physics, University of Silesia, Uniwersytecka 4,
     PL-40007 Katowice, Poland \\
${}^d$Service de Physique Th\'eorique, CEA--Saclay,
          F--91191 Gif-sur-Yvette cedex, France\\
${}^e$Department of Physics, Pennsylvania State University,
           University Park, PA 16802, USA\\
${}^f$Nuclear Physics Institute of Moscow State University,
Moscow 119992, Russia
}

\begin{abstract}
We confirm by explicit computation the conjectured all-orders
iteration of planar maximally supersymmetric $\NeqFour$ Yang-Mills
theory in the nontrivial case of five-point two-loop amplitudes.  We
compute the required unitarity cuts of the integrand and evaluate the
resulting integrals numerically using a Mellin--Barnes representation
and the automated package of ref.~\cite{Czakon}.  This confirmation of
the iteration relation provides further evidence suggesting that
$\NeqFour$ gauge theory
is solvable.
\end{abstract}

\pacs{11.15.Bt, 11.25.Db, 11.25.Tq, 11.55.Bq, 12.38.Bx \hspace{1cm}}

\maketitle

In his seminal work dating to the infancy of asymptotic freedom,
't~Hooft~\cite{tHooft} gave hope of solving quantum chromodynamics
(QCD) in the so-called planar limit, when the number of colors is
taken to be large.  While this hope for ordinary QCD has not yet been
realized, the Maldacena conjecture~\cite{Maldacena} has brought it
closer for four-dimensional maximally-supersymmetric Yang-Mills theory
(\SYM), by proposing a duality relating it at strong coupling to
type~IIB string theory in five-dimensional anti-de Sitter (AdS) space at
weak coupling.  Heuristically, this suggests that the leading-color
terms of the perturbative series should be resummable, and along with
possible non-perturbative contributions should yield relatively simple
results matching those of weakly-coupled gravity. 

While the Maldacena conjecture does not address directly the
scattering amplitudes of on-shell (massless) quanta, previous
work by Anastasiou, Dixon, and two of the authors~\cite{ABDK}
shows that the basic intuition holds.  That paper presented
a conjecture for an all-orders iterative structure in
dimensionally-regulated 
scattering amplitudes of \SYM.  Dixon and
two of the authors~\cite{BDS} fleshed out this structure
for maximally helicity-violating (MHV) amplitudes.
Witten's proposal~\cite{WittenTopologicalString} of a
{\it weak--weak\/} duality between \SYM\
scattering amplitudes and a twistor string theory provides further
indications of new structures underlying the simplicity of both \SYM{}
and string theory in AdS space at strong world sheet coupling.

Ref.~\cite{ABDK} verified the iteration conjecture explicitly for the
two-loop four-point function (a second verification was given in
ref.~\cite{HiddenBeauty}), and ref.~\cite{BDS} did so for the
three-loop four-point amplitude.  Furthermore, the computation of the
two-loop splitting amplitude in ref.~\cite{ABDK}, its own iteration
relation, and consideration of limits as momenta become collinear
shows that were the conjecture to hold for the {\it five\/}-point
two-loop amplitude, it would almost certainly hold for all MHV
two-loop amplitudes.  The step from four-point to five-point
amplitudes is non-trivial, because at five points, functions that are
not detectable in the collinear limits appear~\cite{NeqFourOneLoop}.

An important step in closing this gap has recently been taken by
Cachazo, Spradlin and Volovich~\cite{CSV}.  They confirmed the
conjecture for the terms in the two-loop five-point amplitude even
under parity, using an earlier guess for the integrand~\cite{BRYFive}.
In this Letter, we will complete the task.  We compute the integrand
using the unitarity method~\cite{NeqFourOneLoop,Fusing,BRY},
confirming the form of ref.~\cite{BRYFive} for the parity-even terms,
and providing the correct form for the parity-odd ones.  We then
integrate numerically at random kinematic points, using the \MB{}
integration package~\cite{Czakon}, to show that the conjecture holds
for both parity-even and -odd terms.  We also remark that the
`additional iterative structure' of ref.~\cite{CSV} follows from the
one of ref.~\cite{ABDK} by setting odd parity terms to zero on both
sides of the iteration formula.

The unitarity method~\cite{NeqFourOneLoop,Fusing,TwoloopSplitting} has
proven powerful for computing scattering amplitudes of
phenomenological and theoretical interest out of reach using
conventional Feynman diagrammatic methods.
Improvements~\cite{BCFUnitarity} have followed from use 
of complex momenta~\cite{WittenTopologicalString}.

Perturbative amplitudes in four-dimensional massless gauge theories
contain infrared singularities.  These are well
understood~\cite{CataniDiv} in \SYM\ and are a subset of the ones
appearing in QCD.  As in perturbative QCD, the $S$-matrix under
discussion here is not the textbook one for the `true' asymptotic
states of the four-dimensional theory, but rather for states with
definite parton number. As in QCD, a summation over degenerate states
would be required to obtain finite results for scattering~\cite{KLN}.
We regulate these divergences in a supersymmetry-preserving fashion
using the four-dimensional helicity
(FDH)~\cite{FDH} variant of dimensional regularization, with
$D=4-2\e$.  (This scheme is a close relative of Siegel's dimensional
reduction~\cite{Siegel}.) 

We write the leading-color contributions to the $L$-loop $SU(N_c)$
gauge-theory $n$-point amplitudes as,
\ba
{\cal A}_n^{(L)} &=& g^{n-2}
 \biggl[ { 2 e^{- \gamma \e} g^2 N_c \over (4\pi)^{2-\e} } \biggr]^{L}
 \sum_{\rho}
\Tr( T^{a_{\rho(1)}} 
   \ldots T^{a_{\rho(n)}} )
\nonumber \\&& \null \hskip 10mm\times
               A_n^{(L)}(\rho(1), \rho(2), \ldots, \rho(n))\,,
\ea
where $\gm$ is Euler's constant, 
the sum is over non-cyclic permutations of the external legs.
We have suppressed the momenta and helicities $k_i$ and
$\lambda_i$, leaving only the index $i$ as a label.  This
decomposition holds for all particles in the gauge super-multiplet
as all are in the adjoint representation.  We will find
it convenient to scale out the tree amplitude, defining
$M_n^{(L)}(\e) \equiv A_n^{(L)}/A_n^{(0)}$.

At two loops the iteration conjecture expresses
$n$-point amplitudes entirely in terms of one-loop
amplitudes and a set of constants~\cite{TwoloopSplitting}.  For MHV
amplitudes up to $\Ord(\e^0)$,
\be
M_n^{\twoloop}(\e)
= {1 \over 2} \bigl(M_n^{\oneloop}(\e) \bigr)^2
 + f^\twoloop(\e) \, M_n^{\oneloop}(2\e) + C^{(2)}\,,
\label{TwoloopOneLoop}
\ee
where
$f^\twoloop(\e) = - (\zeta_2 + \zeta_3 \e + \zeta_4 \e^2 + \cdots)$,
and $C^{(2)} = - \zeta_2^2/2$.  Ref.~\cite{BDS} provides analogous
equations for higher-loop MHV amplitudes.  Subtracting out the known
infrared divergences~\cite{CataniDiv} provides an all-loop form for
the finite remainder, expressed in terms of the one-loop finite
remainder and two constants, one of which is an anomalous dimension.
A conjecture for the required anomalous dimension was
very recently presented~\cite{StaudacherSoftAnomDim}, based on a
proposed all-loop Bethe Ansatz~\cite{BeisertStaudacherAllLoop}. 
It is rather
interesting that this anomalous dimension corresponds to one of the
terms appearing in the QCD one~\cite{KLOV}.

\begin{figure}[t]
\centerline{\epsfxsize 3 truein \epsfbox{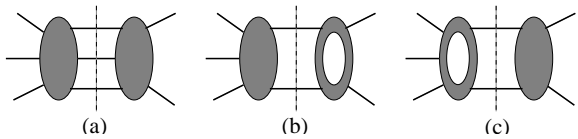}}
\caption[a]{\small The three- and two-particle 
cuts of the five-point amplitude.}
\label{TwoThreeCutFigure}
\end{figure}

To check whether the iteration relation holds in the critical 
five-point case,
 we have evaluated a set of cuts sufficient to determine the
five-gluon integrand completely.  These include the three-particle
cuts depicted in \fig{TwoThreeCutFigure}{a}
as well as the contributions to the two particle cuts from
\fig{TwoThreeCutFigure}{b}.  The three-particle cuts on their own determine
all integral functions, except for those which are simple products
of one-loop integrals.  The
two-particle cuts rule out the latter (double cuts suffice).  

The use of a dimensional regulator involves an
analytic continuation of the loop momenta to $D$ dimensions. At one
loop, the discrepancy between treating loop momenta
in four or $D$ dimensions does not modify the amplitudes of a
supersymmetric gauge theory through $\Ord(\ep^0)$.  No such
proof exists for higher loops.  Thus, to ensure that no contributions are
dropped, we compute the unitarity cuts in $D$
dimensions~\cite{DDimUnitarity}.  This does complicate the analysis, because
standard helicity states can no longer be used as the intermediate
states.  We can avoid some of the additional complexity
by considering instead the $D=10,\,
{\cal N}=1$ super-Yang-Mills theory. When compactified on a torus to
$D=4-2\ep$ dimensions this is equivalent to dimensionally-regulated
\SYM\ in the FDH scheme.

\begin{figure}[t]
\centerline{\epsfxsize 3. truein \epsfbox{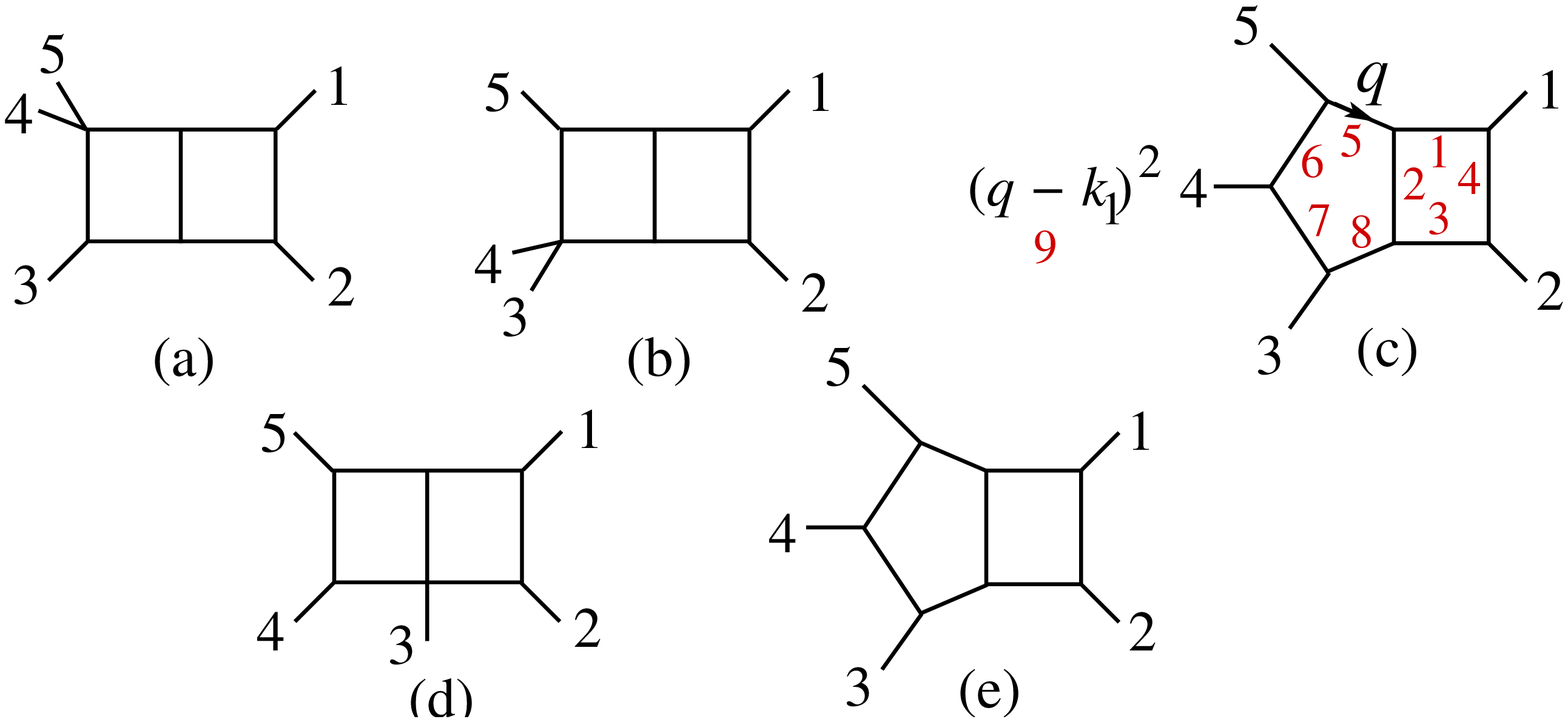}}
\caption[a]{\small The two-loop integrals appearing in the
five-point amplitude, with all external momenta flowing outwards.
The normalization is
as given in \eqn{Penta2MB}, and the numerical labels on the internal
propagators in (c) specify the arbitrary powers $a_i$.  The prefactor in (c)
is understood to be inserted in the numerator with power $-a_9$;
in \eqn{TwoloopIntegrand}, $-a_9 = 1$. }
\label{TwoloopIntegralsFigure}
\end{figure}

After reducing all tensor integrals we obtain an expression for the
amplitude in terms of the integrals shown
in~\fig{TwoloopIntegralsFigure}. The color-ordered amplitude with
four-dimensional external momenta is given by a sum over the cyclic
permutations of those momenta,
\begin{widetext}
\ba
M_5^\twoloop(\ep) &=& {1\over 8} \sum_{\rm cyclic} \Biggl\{
           s_{12}^2 s_{23} \,I^{\twoloop}_{\rm (a)}(\ep)
         + s_{12}^2 s_{15} \,I^{\twoloop}_{\rm (b)}(\ep)
         + s_{12} s_{34} s_{45}\, I^{\twoloop}_{\rm (c)}(\ep) \nn \\
&& \null + R
  \biggl[
       2\, I^{\twoloop}_{\rm (d)}(\ep) - 2\, s_{12} \,
                                           I^{\twoloop}_{\rm (e)}(\ep)
      + { s_{12} \over s_{34} s_{45} }
         \biggl({\delta_{-++}\over s_{23}} \, I^{\twoloop}_{\rm (b)}(\ep)
       - {\delta_{-+-} \over s_{51} } \, I^{\twoloop}_{\rm (a)}(\ep)  \biggl)
       + {\delta_{+-+} \over
          s_{23} s_{51} }\, I^{\twoloop}_{\rm (c)}(\ep) \biggr] \Biggl\}   \,.
\hskip .3 cm
\label{TwoloopIntegrand}
\ea
\end{widetext}
\def\sliver{\mskip2mu}
Here, $s_{ij}=(k_i+k_j)^2$, $R=\vareps_{1234}
        {s_{12} s_{23} s_{34} s_{45} s_{51}/G_{1234}}$,
\ba
\delta_{abc} = s_{12} s_{51} +a\sliver s_{12} s_{23} + b\sliver
                  s_{23} s_{34} - s_{51} s_{45} + c\sliver s_{34} s_{45}
\sliver, \nn\\
\varepsilon_{1234} = 4 i \,
 \varepsilon_{\mu\nu\rho\sigma} k_1^\mu k_2^\nu k_3^\rho k_4^\sigma
 = \tr[\gamma_5 \ksl_1 \ksl_2 \ksl_3 \ksl_4] \,,~~~~~~~~~~~~
\ea
and $G_{1234} = \det(s_{ij}) \,, ~ (i,j = 1, \ldots, 4)$.
(In $\delta$, $a,b,c = \pm1$.)
The terms lacking a factor of
$\vareps_{1234}$ are even under parity, while those
with such a factor are odd.  The even terms match
the guess originally given in ref.~\cite{BRYFive}, but the odd
terms differ (the odd terms in
ref.~\cite{BRYFive} do match the four-dimensional double
two-particle cuts).

%
\begin{figure}[t]
\centerline{\epsfxsize 1.4 truein \epsfbox{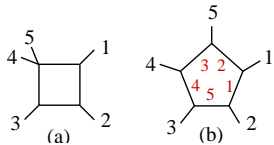}}
\caption[a]{\small The one-loop integrals required to all orders
in $\e$ for the one-loop five-point amplitude.  The normalization is
as given in \eqn{Penta1MB}, and the numerical labels on the internal
propagators in (b) specify the arbitrary powers $a_i$.}
\label{OneloopIntegralsFigure}
\end{figure}

Because of the $1/\ep^2$ infrared singularity in one-loop amplitudes,
and because these appear squared in the iteration relation, we need
expressions valid through $\Ord(\ep^2)$.  A representation of the
one-loop five-point amplitude, extending ref.~\cite{NeqFourOneLoop} to
all orders in $\e$, may be found in ref.~\cite{DimShift},
\be
 M^\oneloop(\ep) = -{1\over 4} \! \sum_{\rm cyclic} \hskip -.1 cm 
                          s_{12} s_{23} I^{\oneloop}_{\rm (a)} (\ep)
   - {\ep \over 2} \, \vareps_{1234} I^{\oneloop 6-2\ep}_{\rm (b)}(\ep) \,,
\label{OneloopAmplitude}
\ee
in terms of the integrals of~\fig{OneloopIntegralsFigure}. As indicated by
the superscript, the second integral (fig.~\ref{OneloopIntegralsFigure}b)
is to be evaluated in $6-2\ep$ dimensions.  In $D=6$ it
is completely finite, but because it appears multiplied by an
infrared-singular integral in \eqn{TwoloopOneLoop} we need its value through
$\Ord(\ep)$.

To obtain Laurent expansions in $\ep$ for our integrals
we use the Mellin--Barnes (MB) technique, 
successfully applied in numerous calculations
(see, e.g., refs.~\cite{MB1,MB2,Threeb,BDS,S2} and
chap.~4 of ref.~\cite{S4}).
It relies on the identity
\ba
\frac{1}{(X+Y)^{\lm}} =
\int_{\beta - i  \infty}^{\beta + i  \infty}
\frac{Y^z}{X^{\lm+z}} \frac{\Gm(\lm+z) \Gm(-z)}{\Gm(\lm)}
\frac{\dd z}{2\pi  i} \,,
\ea
where $-{\rm Re}\, \lm <\beta <0$.  This basically replaces a sum over terms
raised to some power with a product of factors.

The box function in fig.~\ref{OneloopIntegralsFigure}a was given
to all orders in $\e$ in terms of a hypergeometric function
in ref.~\cite{OneloopPentagon}. Here we
need its value through $\Ord(\ep^2)$.
Evaluating the pentagon in \fig{OneloopIntegralsFigure}b
with arbitrary powers of propagators also allows a parallel evaluation
of this integral to the required order.

\def\aa{{A}}
The derivation of a fourfold MB representation for the one-loop
pentagon diagram is straightforward, after Feynman parameterizing,
\def\spacer{\hskip 10mm}
\def\spacerB{\hskip 7mm}
\begin{widetext}
\ba
&& \hskip -.3 cm 
P^{(1)}(a_1,\ldots,a_{5};s_{12},\ldots,s_{51};\ep)
= -i\, e^{\gamma \ep} (4 \pi)^{D/2}  \int {d^D\ell \over (2\pi)^D} \,
{1\over [\ell^2]^{a_1} [(\ell+k_1)^2]^{a_2}
[(\ell+K_{15})^2]^{a_3} [(\ell-K_{23})^2]^{a_4} [(\ell-k_{2})^2]^{a_5}}
\nn\\
&&\spacer= \frac{e^{\gm \ep} (-1)^\aa }{
 \Gm(4-\aa -2\ep)}
\frac{1}{\prod_{j=1}^5\Gm(a_j)}
\int_{-i\infty}^{+i\infty}
 \prod_{j=1}^4  \Gm(-z_j) \frac{\dd z_j}{2\pi i} 
\frac{(-s_{45})^{z_1} (-s_{34})^{z_2} (-s_{23})^{z_3}
(-s_{12})^{z_4}}{(-s_{15})^{\aa + \ep - 2 + z_{1234}}}
\Gm(a_2 + z_{14})
\nn \\
&&\spacer\spacerB\null\times
\Gm(\aa + \ep - 2 + z_{1234})
\Gm(2 - \ep - a_{2345} - z_{124})
\Gm(2 - \ep - a_{1245} - z_{134})
 \Gm(a_4 + z_{13}) \Gm(a_5 + z_{24})
\label{Penta1MB}
\ea
\end{widetext}
where $K_{ij} = k_i+k_j$, $a_{2345}=a_2+a_3+a_4+a_5, \aa=\sum a_i,
z_{124}=z_1+z_2+z_4$, etc.  We have allowed for arbitrary powers of
propagators so that we can obtain all one-loop integrals. Taking $a_5
\rightarrow 0$, with other $a_i=1$, gives the box integral
$I^{\oneloop}_{\rm (a)}$ in \fig{OneloopIntegralsFigure}a.  Setting
all $a_i = 1$ and shifting all terms except the $e^{\gamma \e}$
prefactor by $\ep \rightarrow \ep - 1$ yields the $D=6-2\ep$ pentagon,
$I^{\oneloop 6-2\ep}_{\rm (b)}$, corresponding to
\fig{OneloopIntegralsFigure}b.

The various two-loop pentabox integrals have a sevenfold
MB representation obtained by inserting a threefold MB
representation for a two-mass double box into
\eqn{Penta1MB},
\def\spacerC{\hskip 3mm}
\begin{widetext}
\ba
&& \hskip -.6 cm 
P^{(2)}(\{a_i\};\{s_{ij}\};\ep)
=\frac{ e^{2\gm \ep} (-1)^A}{
\prod_{j=1,2,3,4,6,7}\Gm(a_j) \Gm(4-a_{1234} -2\ep)}
\int_{-i\infty}^{+i\infty}
\prod_{j=1}^7  \Gm(-z_j) \frac{\dd z_j}{2\pi i}
\nn\\ &&\spacer\spacerC\null\times
\frac{ (-s_{45})^{z_1} (-s_{12})^{2 - a_{1234} - \ep + z_4 - z_{567}}
(-s_{23})^{z_3} (-s_{34})^{z_2}}
{(-s_{15})^{a_{56789} + \ep-2 + z_{1234} - z_{567}} }
\frac{\Gm(a_7 + z_{13})
\Gm(a_5 + z_{14} - z_5) \Gm(a_8 + z_{24} - z_6)}
{\Gm(a_5 - z_5) \Gm(a_8 - z_6) \Gm(a_9 - z_7)}
\nn\\ &&\spacer\spacerC \null \times
\frac{\Gm(2 - \ep- a_{5678}  - z_{124} + z_{56})
\Gm(a_4 + z_7)\Gm(a_2 + z_{567})\Gm(a_{56789} + \ep-2 + z_{1234} - z_{567})}
{\Gm(4 - 2 \ep- a_{56789}  + z_{567})}
\label{Penta2MB}
\\ &&\spacer\spacerC \null \times
\Gm(2  - \ep - a_{124}- z_{57})
\Gm(2 - \ep- a_{234}  - z_{67})
\Gm(a_{1234} + \ep-2 + z_{567})
\Gm(2- \ep - a_{5789}  - z_{134} + z_{567})
\,.\nn
\ea
\end{widetext}
The limit $a_6 \rightarrow  0$ or $a_7 \rightarrow  0$
with $a_9=0$ and the other $a_i=1$ yields
the double box with
one massive leg (\fig{TwoloopIntegralsFigure}a and b) in
agreement with 
ref.~\cite{S2,LGR2}. Moreover,
$P^{(2)}(1,\ldots,1,-1),\,P^{(2)}(1,\ldots,1,0)\;{\rm and}\;
P^{(2)}(1,\ldots1,0,0)$ yield the integrals in \fig{TwoloopIntegralsFigure}c, 
\fig{TwoloopIntegralsFigure}e and \fig{TwoloopIntegralsFigure}d,
respectively.

An essential step in the use of the MB technique
is the resolution of singularities in $\ep$ or zeros that
appear as $a_i \rightarrow 0$.  There are two strategies for doing
this \cite{MB1,MB2}.  Quite
recently, the second strategy was formulated algorithmically
\cite{AnDa,Czakon} and implemented in the \MB{} package
\cite{Czakon}.  It produces code that allows the integrals
to be evaluated numerically to reasonably high accuracy.

The even terms in \eqn{TwoloopIntegrand} were recently evaluated
in ref.~\cite{CSV} using the {\tt MB\/} package~\cite{Czakon} along
with the guess of
ref.~\cite{BRYFive}.  Those authors observed that an iterative
structure holds for the parity-even terms alone.
We may observe that this structure is not independent of
the complete iteration formula \eqn{TwoloopOneLoop}: use
the results for the one-loop five-point amplitude in
\eqn{OneloopAmplitude}, set the odd terms to zero and use
the fact that the one-loop MHV amplitudes have even parity
through ${\cal O}(\epsilon^0)$.
At higher loops, we do not expect a clean
separation between `even' and `odd' terms, as non-vanishing
terms of the form
$\vareps^2_{1234}$ will arise.  These are even under parity.

We have evaluated
all the two-loop integrals in \fig{TwoloopIntegralsFigure}
through $\Ord(\ep^0)$ and the one-loop integrals in
\fig{OneloopIntegralsFigure} through $\Ord(\ep^2)$ using
the representations in \eqns{Penta1MB}{Penta2MB}.
 We have checked to a
numerical accuracy of five significant digits at three independent
kinematic points that the iteration formula~(\ref{TwoloopOneLoop}) is
indeed correct for the complete amplitude.
This is a crucial check on the conjecture because
the parity-odd terms in the five-point amplitude are precisely
the ones which are not constrained by collinear factorization
onto four-point amplitudes.

The calculation presented here makes a nontrivial addition to the
existing body of evidence for the iteration
conjecture~\cite{ABDK,BDS}.  The conjecture itself gives us good
reasons to believe that \SYM\ is solvable.  Within the context of the
planar perturbative expansion, this would imply the resummability of the
series.  Parallel developments in uncovering the integrable structure
of the theory (see e.g. refs.~\cite{MZ,BeisertStaudacherAllLoop}) also
lend credence to this belief.

\smallskip
We thank Lance Dixon for many helpful discussions, and participation
in early stages of the collaboration.  This research was supported by
the US DOE under contracts DE--FG03--91ER40662, by the {\it Direction
des Sciences de la Mati\`ere\/} of the {\it Commissariat \`a l'Energie
Atomique\/} of France, and by the Russian Foundation for Basic
Research through grant 05--02--17645. The work of MC was supported by the
Sofja Kovalevskaja Award of the Alexander von Humboldt Foundation
sponsored by the German Federal Ministry of Education and Research.
DAK and VAS also acknowledge the support of the ECO-NET program of the
{\sc Egide} under grant 12516NC.

${}$

\end{document}